\documentclass[conference]{IEEEtran}
\IEEEoverridecommandlockouts
\usepackage{cite}
\usepackage{amsmath,amssymb,amsfonts}
\usepackage{algorithmic}
\usepackage{graphicx}
\usepackage{textcomp}
\usepackage{xcolor}
\usepackage{epstopdf}
\usepackage{booktabs}
\usepackage {mathtools}
\usepackage{multirow}
\def\BibTeX{{\rm B\kern-.05em{\sc i\kern-.025em b}\kern-.08em
    T\kern-.1667em\lower.7ex\hbox{E}\kern-.125emX}}
\begin{document}

\title{An Evidential Reasoning Based Approach to Building Node Selection Criterion for Network Reduction\\
}

\author{\IEEEauthorblockN{Bin Huang, Jiayong Li, and Jianhui Wang*}
	\IEEEauthorblockA{\textit{Dept. of Electrical and Computer Engineering} \\
		\textit{Southern Methodist University}\\
		Dallas, Texas \\
		jianhui@smu.edu}}

\maketitle

\begin{abstract}
A reasonable node selection criterion (NSC) is crucial for the network reduction in power systems. In contrast to the previous works that only consider structure property, this paper proposes a comprehensive and quantitative NSC considering both structural and electrical properties. The proposed NSC is developed by employing the evidential reasoning approach, in which the quasi-one-hot encoding is used to determine the evaluation grades of different criteria or attributes. Then, different criteria are combined through the multi-evidence reasoning. Eventually, the utility evaluation is used to derive the quantitative NSC. Besides, the ER can be readily extended to multiple criteria while considering the uncertainty in the evaluation process simultaneously. The reduced models with higher accuracy can be built by combining the proposed NSC with the existing model reduction algorithms. The case studies on a 30-node power grid substantiate the practicality of the proposed NSC.
\end{abstract}


\section{Introduction}
Network reduction is crucial to the operation and analysis of the interconnected power gird\cite{tinney1987adaptive}. The goal of the network reduction or equivalent is to represent the original power grid with a smaller equivalent model, which can enhance the computation speed, alleviate the memory requirement, and improve the analysis efficiency. For example, it is reported in \cite{pecenak2017multiphase} that the time for evaluating the impact of the integration of photovoltaic on a 621-bus feeder was reduced by up to 96\% with the help of network reduction. Moreover, the control centers of different areas of power grids may be unwilling to share their private data such as detailed models with others due to independent operation. Network reduction is one of the handy techniques to tackle this issue. For instance, \cite{xu2017calculation} employs network reduction techniques to calculate the total transfer capability of the multi-area power grids.    

In general, the reduction or equivalence approaches can be categorized into the dynamic equivalent methods and static equivalent methods (SEMs). The former considers the dynamic and transient characteristics of the power grids while the latter focus on static analysis and operation. We focus on the latter in this paper. Researchers have proposed different SEMs to construct equivalent models, among which Ward-type\cite{tinney1987adaptive}, radial equivalent independent circuits (REI)\cite{irisarri1979real}, bus aggregation\cite{oh2012aggregation,shi2014novel} and measurement-based methods\cite{yu2017optimal,samadi2015static} are adopted widely. Generally speaking, the procedure of the SEMs can be summarized into four steps: 1. differentiate the retained nodes and eliminated nodes; 2. construct the reduced model with circuit components such as the equivalent lines, equivalent generators and equivalent loads; 3. formulate the specific mathematical models (nonlinear equations or optimization model) to determine the reduced model parameters; 4. access the accuracy of the reduced model.

Differentiating retained nodes and eliminated nodes plays a crucial role in the network reduction. For some specific applications such as the interconnected power grid analysis \cite{tinney1987adaptive,yu2017optimal}, people tend to be interested in only one subsystem. In this scenario, the retained nodes are highly aggregated in one subsystem, in which a retained area can be formed. The retained area and the eliminated area are coupled through only a few boundary nodes, which means the topology and electrical connections between the retained and eliminated area are weak. However, in some issues such as optimal generation investment planning\cite{shi2012optimal,shi2014novel} and distributed energy resource planning\cite{pecenak2017multiphase}, the retained nodes are scattered across the whole system. In \cite{shi2012optimal}, the criteria to select the retained nodes is transmission congestion. Those lines with the highest degree of congestion and their corresponding end nodes are preserved. In \cite{shi2014novel}, the network to be reduced is divided into different modules by utilizing the power transfer distribution factors (PTDFs). The nodes within a module aggregate into one artificial node on a module level. \cite{pecenak2017multiphase} retains the critical nodes selected by users and identify the rest of them using topology analysis. The node selection strategy in \cite{pecenak2017multiphase} is tailored specially for the distribution systems with radial topology\cite{zhang2019interval}, thus it can not be applied in the transmission systems directly. Overall, these node selection strategies are either experience-dependent or they only considered one of the critical characteristics at a time (topology, congestion, etc.). Thus, a comprehensive and judicious node selection criterion (NSC) is imperative for the model reduction of power grids.

Complex networks are known to be a powerful and flexible representation formalism. The complex networks can represent not only the feature values of individuals but also connection relationships among different individuals. Based on the matching of bus-node and branch-edge between the power grid and the complex networks, numerous researches have attempted to analyze the power grids from the perspectives of complex networks\cite{bompard2011structural,arianos2009power}.
Leveraging the complex networks theory can unveil the essential topology properties such as betweenness and closeness of power grids. By combining the electrical features of power grids and topology properties derived from the complex network, the synthetic properties can be identified. These properties are useful in network reduction problems. Not only can they guide the strategy of retaining nodes, but also they provide a new angle for evaluating the accuracy of the equivalent models.

In this paper, a comprehensive and quantitative node selection criterion for network reduction is proposed for the first time. The criterion is derived through the combination of dual criteria using the evidential reasoning approach (ER), which is generic and can be extended to involve multiple criteria. In particular, it takes both the electrical features and topology into account to fully exploit the system information. The reduced model derived from the proposed criterion is closer to the full model compared with those from a single criterion on various structural properties. The feasibility of the proposed criterion is verified through the case studies upon a 30-node power grid. Furthermore, the proposed method to derive the criterion can preprocess nodal features in graph learning. Thus its application has a promising future in the field of machine learning which deals with graph data, such as network representation learning \cite{zhang2018network} and graph convolutional networks\cite{hen2019fault}.
\vspace{-0.2cm}
\section{Nodes Importance Criteria}\label{sec1}
\vspace{-0.1cm}
In this section, two criteria for evaluating the importance of nodes in power grids are investigated. One is the extended betweenness, which is analogous to the betweenness centrality in complex networks. The other is net-ability, which is analogous to the global efficiency in complex networks. In addition to the abstract topological properties, these two criteria consider the detailed physical characteristics of the electrical networks. The nodal rankings of the electrical networks obtained from these two criteria provide straightforward guidelines to NSC: the nodes with high ranking can be reserved, whereas those with low ranking can be eliminated.
\vspace{-0.1cm}
\subsection{Extended Betweenness}
From the perspective of complex networks, a power grid can be represented as a graph $G=\{V,E,W\}$, where $V$ is the node set, $E$ is the line set, and $W$ is the weight set for branches.

Betweenness centrality is a metric which can assess the importance of nodes by quantifying the role of nodes in the information exchange of complex networks. Nodal betweenness is defined by the probability of the node exists in the shortest paths of all node pairs of the graph. The mathematical formulation of betweenness centrality can be expressed as \cite{brandes2001faster}:
\begin{equation}\label{eq_abs_between}
B_{0}(v) = \sum_{v \not= f \not= t \in V } \frac{\varphi_{ft}(v)}{\varphi_{ft}}
\end{equation}
where subscripts $f$ and $t$ denote the node pair $\{f,t\}$, $\sum_{f \not= t} \varphi_{ft}$ denote the number of shortest paths of the node pairs, and $\varphi_{st}(v)$ is the number of the shortest paths that go through node $v$.

However, (\ref{eq_abs_between}) is not a pragmatic criterion to determine the importance of nodes in power grids since it ignores the physical features of power grids. To address this issue, an extended betweenness criterion was developed in \cite{bompard2011structural}. One feature of this criterion is that it only considers the node pairs between generator nodes and load nodes. The definition of the extended betweenness is based on PTDFs, which represent the linearized relationship between the power injections of node pairs and the branch flows. The extended betweenness is formulated as:
\begin{equation}\label{eq_ele_between}
\vspace{-0.2cm}
\hat{B}(v) = \frac{1}{2} \sum_{g \in Gen} \sum_{d \in D} \kappa_{g}^{d} \sum_{l \in L^{v}} |\tau^{gd}_{l}|, v \not= g \not= d \in V
\end{equation}
where $Gen$ is the generator nodes set, $D$ is the loads nodes set, $\tau^{gd}_{l}$ is the PTDF w.r.t node pair $\{g,d\}$ and line $l$, $L^{v}$ is the set of lines connecting node $v$ directly. $\kappa_{g}^{d}$ is defined as the power transmission capacity, which is formulated as:
\begin{equation}\label{eq_trans_cap}
\kappa_{g}^{d} = \min_{l \in L}(\frac{P_{l}^{\max}}{|\tau^{gd}_{l}|})
\end{equation}
where $P_{l}^{\max}$ is the capacity of line $l$.

More specifically, $\kappa_{g}^{d}$ represents the maximum transmission power of the node pair $\{g,d\}$ because at least one line in the system will reach its capacity when the power transfer between the node pair $\{g,d\}$ is $\kappa_{g}^{d}$. Thus, (\ref{eq_ele_between}) can be interpreted as the total power flowing through the node $v$ when all nodes pairs $\{g,d\}$ in system transfer the power with the amount of $\kappa_{g}^{d}$. 

\subsection{Net-ability}
In complex networks theory, the global efficiency of a network is a metric measuring how efficiently it exchanges information. The formulation of the global efficiency of complex networks is:
\begin{equation}\label{eq_abs_Eglo}
\vspace{-0.2cm}
E_{\text{global}} = \frac{1}{N_{V}(N_{V}-1)} \sum_{f \not= t \in V} {d_{ft}^{-1}}
\end{equation}
where $d_{ft}$ is the shortest path length between node $f$ and $t$ and $N_{V}$ is the number of nodes.

\cite{arianos2009power} defines the efficiency of electrical networks by replacing the path length with the electrical distance. The electrical distance is determined by the impedance matrix of the system and the detailed calculation can be referred to \cite{arianos2009power}. Therefore, the global efficiency of power grids can be represented as:
\begin{equation}\label{eq_ele_Eglo}
\vspace{-0.2cm}
A_{\text{global}} = \frac{1}{N_{Gen}N_{D}} \sum_{g \in Gen} \sum_{d \in D} \frac{\kappa_{g}^{d}}{Z_{g}^{d}}
\end{equation}
where $N_{Gen}$ and $N_{D}$ are the number of generator and load nodes, respectively. $Z_{g}^{d}$ is the equivalent impedance of the node pair $\{g,d\}$.

Dropping nodes from the system one by one and evaluating the impact they cause is a useful way to measure the importance of nodes. Note that the removal of nodes will lead to the removal of corresponding lines as well. In this manner, the net-ability of node $v$ is defined as:
\begin{equation}\label{eq_netability}
\Delta A_{v} = \frac{|A_{\text{global}}^{\text{ori}} - A_{\text{global}}^{\text{re},v}|}{A_{\text{global}}^{\text{ori}}}
\end{equation}
where $A_{\text{global}}^{\text{ori}}$ and $A_{\text{global}}^{\text{re},v}$ are the global efficiency of the original and the reduced model without node $v$, respectively.
	 

\section{Comprehensive Criterion Based on Evidential Reasoning}\label{sec:3}
According to the simulation studies presented in \cite{bompard2011structural}, there is a discrepancy between the node rankings obtained from the extended betweenness and net-ability. Therefore, a comprehensive ranking criterion that can take into account the multiple evaluation criteria simultaneously is of great interest. 

Evidential reasoning (ER) was developed in \cite{yang2002evidential} to tackle the multi-attribute decision making (MADM) under uncertainty. The ER can adequately consider multiple criteria and their uncertainty. In this section, a comprehensive criterion is developed using the ER. The ER method is constituted by three steps: multi-attribute analysis, multi-evidence reasoning, and utility evaluation, which are described in detail below.

\subsection{Multi-attribute analysis}
The net-ability and extended betweenness are regarded as two attributes of ER in this paper.
For assessing the state of each attribute, the evaluation grades are defined as:
\begin{equation}\label{grades}
S = \{S_{1},S_{2},\dots,S_{n},\dots,S_{N}\}
\end{equation}
In our derivation, the elements in $S$ is constituted by the index of nodal ranking, thus $N$ is identical to $N_{V}$. Note that the descending ranking is adopted in this paper. 

Afterwards, for each node $x$, we can determine the degree of belief of each attribute on $S$:
\begin{equation}\label{eq8}
\Psi_{x}(a_{j}) = \{(S_{n},\delta_{n,j}, n = 1,\dots,N)\} \quad j=1,\dots,\Lambda
\end{equation}
This equation indicates the attribute $a_{j}$ of node $x$ has $\delta_{n,j}$ degree of belief on $H_{n}$. Note that $\delta_{n,j} \geq 0 \enspace \text{and} \enspace \sum_{n=1}^{N}\delta_{n,j} \leq1$. $\Lambda$ is the number of attributes and is set to 2 (the number of criteria) in this case. 

Considering the nature of the ranking problem, the assignment of $\delta_{n,j}$ adopts the one-hot encoding. Specifically, in this case, $\delta_{n,j}$ is assigned to 0.9 if the node ranks $n$th in attribute $j$ and the rest grades of attribute $j$ is set to 0. Setting the degree of belief to 0.9 rather than 1 represents the uncertainty of the criterion. The uncertainty means one specific criterion is not able to include the overall information of the system. In practical application, assessors can adjust these parameters to represent the degree of comprehensiveness of criterion. For illustration, Table \ref{tab3node1} displays the basic assessment table of a three-node system, in which the rank of nodes on the extended betweenness and net-ability are \{3,2,1\} and \{1,2,3\}, respectively.

\subsection{Multi-evidence reasoning}
Afterward, the function of multi-evidence reasoning is combining multiple attributes to acquire a synthetic attribute. 

Let $\omega_{j}$ denote the weight of attribute $j$. Define $e_{n,j}$ as the basic probability mass, which represent the weighted degree of belief and can be calculated by $e_{n, j}=\omega_{j} \delta_{n, j}$. The remaining probability mass $e_{S, j}$ represent the uncertainty in evaluation grade, and is decomposed into $\bar{e}_{S, j}=1-\omega_{j}$ and $\tilde{e}_{S, j}=\omega_{j}(1-\sum_{n=1}^{N} \delta_{n, j})$. 

The aggregation of attributes is conducted in a recursive manner, which starts from the first attribute and ends in the last attribute. Denote $e_{n, I(j)}$, $\tilde{e}_{S, I(j)}$, and $\bar{e}_{S, I(j)}$ be the basic probability mass, the first remaining probability mass, and the second remaining probability mass after combining the first j (from 1 to j) attribute, respectively. The recursive processes are developed as \cite{yang2002evidential,wu2019large}:
\vspace{-0.2cm}
\begin{equation}
\begin{aligned}
\left\{H_{n}\right\}: e_{n, I(j+1)}=K_{I(j+1)}e_{n, I(j)} e_{n, j+1}\\
+K_{I(j+1)}(e_{S, I(j)} e_{n, j+1}+e_{n, I(j)}, e_{S, j+1})
\end{aligned}
\end{equation}
\vspace{-0.2cm}
\begin{equation}
e_{S, I(j)}=\tilde{e}_{S, I(j)} + \bar{e}_{S, I(j)}
\end{equation}
\vspace{-0.2cm}
\begin{equation}
\begin{aligned}
\{S\}: \tilde{e}_{S, I(j+1)}=K_{I(j+1)} \tilde{e}_{S, I(j)} \tilde{e}_{S, j+1}\\
+ K_{I(j+1)} ( \bar{e}_{S, I(j)} \tilde{e}_{S, j+1}+\tilde{e}_{n, l(j)} \bar{e}_{S, j+1})
\end{aligned}
\end{equation}
\vspace{-0.2cm}
\begin{equation}
\bar{e}_{S, I(j+1)}=K_{I(j+1)} \bar{e}_{S, I(j)} \bar{e}_{S, j+1}
\end{equation}
where $K_{I(j+1)}$ is the aggregation coefficient and can be calculated by:
\vspace{-0.1cm}
\begin{equation}
\vspace{-0.2cm}
K_{I(j+1)}=(1-\sum_{k=1, k \neq t}^{N} \sum_{t=1}^{N}  e_{t, I(j)} e_{k, j+1})^{-1}
\end{equation}

The attributes-combined degree of belief $\delta_{n}^{\text{com}}$ corresponding to the grade $S_{n}$ is calculated by:
\vspace{-0.1cm}
\begin{equation}
\vspace{-0.2cm}
\left\{S_{n}\right\}: \delta_{n}^{\text{com}}= (1-\bar{e}_{S, I(\Lambda)} )  ^{-1} {e_{n, I(\Lambda)}}
\end{equation}

The uncertain attributes-combined degree of belief $\delta_{S}$ is calculated by:
\vspace{-0.2cm}
\begin{equation}
\vspace{-0.2cm}
\{S\}: \delta_{S}= ({1-\bar{e}_{S, I(\Lambda)}})^{-1} {\tilde{e}_{n, I(\Lambda)}}
\end{equation}

Consequently, $\Psi$ in (\ref{eq8}) can be compacted as the attributes-combined form:
\begin{equation}\label{eq11}
\begin{aligned}
{\boldsymbol{\tilde{\Psi}}_{y}}=\boldsymbol{\Psi}_{y}\left(a_{1} \oplus a_{2} \cdots \oplus a_{j} \oplus \cdots \oplus a_{\Lambda}\right) = \left\{\left(S_{n}, \delta_{n}^{\text{com}}\right) \right\}
\end{aligned}
\end{equation}
where $\oplus$ is the combination operator.

With (\ref{eq11}), we can obtain the attributes-combined evaluation table for the ranking of nodes. Table \ref{tab3node2} displays such a table for the three-node system we discuss above.

\subsection{Utility Evaluation}
It is not straightforward to rank the nodes with the attributes-combined evaluation table. Eventually, utility evaluation is used to determine the ultimate ranking of the nodes. First, we can define the utility function $u$ as:
\begin{equation}
u(H_{n}) = \frac{N-n}{N} \quad n=1,2,\cdots,N
\end{equation}

By assigning the $\delta_{S}$ to the utility function $u(S_{1})$ and $u(S_{N})$, we can obtain the maximum, minimum and average utilities for node $x$:
\begin{equation}
\vspace{-0.2cm}
u_{\min }(x)=\sum_{n=1}^{N-1} \delta_{n} u\left(S_{n}\right)+\left(    \delta_{N}+    \delta_{S}\right) u\left(S_{N}\right)
\end{equation}
\begin{equation}
\vspace{-0.2cm}
u_{\max }(x)=\left(    \delta_{1}+    \delta_{S}\right) u\left(S_{1}\right)+\sum_{n=2}^{N} \delta_{n} u\left(S_{n}\right)
\end{equation}
\begin{equation}
\vspace{-0.2cm}
u_{\mathrm{avg}}(x)=\frac{\left(u_{\max }(x)+u_{\min }(x)\right)}{2}
\end{equation}

Herein average utilities are regarded as the ultimate criterion for ranking the importance of nodes. The development of average utilities here involves extended betweenness, net-ability, and uncertainty. After obtaining the average utility for each node in the system, we can rank the nodes in power grids, thereby determining the retained nodes in the network reduction. ER is a flexible approach. Apart from the net-ability and extended betweenness, other criteria can be integrated into the ER approach readily by modifying the attributes set.

\begin{table}[]
	\caption{An illustrative example of $\Psi(a_{j})$ of three-node system}\label{tab3node1}
	\vspace{-0.4cm}
	\begin{center}
	\begin{tabular}{cccccc}
	\hline
	\multirow{2}{*}{Node} & \multirow{2}{*}{Attribute} & \multicolumn{4}{c}{Evaluation grade} \\ \cline{3-6} 
	&                            & 1     & 2     & 3     & uncertainty  \\ \hline
	\multirow{2}{*}{1}    & Extended betweenness       & 0     & 0     & 0.9   & 0.1          \\  
	& Net-ability                & 0.9   & 0     & 0     & 0.1          \\ 
	\multirow{2}{*}{2}    & Extended betweenness       & 0     & 0.9   & 0     & 0.1          \\  
	& Net-ability                & 0     & 0.9   & 0     & 0.1          \\ 
	\multirow{2}{*}{3}    & Extended betweenness       & 0.9   & 0     & 0     & 0.1          \\  
	& Net-ability                & 0     & 0     & 0.9   & 0.1          \\ \hline
	\end{tabular}
	\end{center}
	\vspace{-0.4cm}
\end{table}

\begin{table}[]
	\caption{An illustrative example of $\tilde{\boldsymbol{\Psi}}$ of three-node system}\label{tab3node2}
	\vspace{-0.4cm}
	\begin{center}
	\begin{tabular}{ccccc}
		\hline
		\multirow{2}{*}{Node} & \multicolumn{4}{c}{Evaluation grade} \\ \cline{2-5} 
		& 1      & 2     & 3     & uncertainty \\ \hline
		1                     & 0.452  & 0     & 0.452 & 0.096       \\ 
		2                     & 0      & 0.93  & 0     & 0.07        \\ 
		3                     & 0.452  & 0     & 0.452 & 0.096       \\ \hline
	\end{tabular}
	\end{center}
	\vspace{-0.4cm}
\end{table}

\section{Network reduction} \label{sec4}
With the comprehensive criterion derived from ER,  the retained nodes and eliminated nodes in power grids can be differentiated. The modified Ward equivalence method developed in \cite{shi2012optimal}  is adopted in this paper to achieve network reduction. Note that the method developed in \cite{shi2012optimal} is a direct current (DC) power flow-based method, which is designed mainly for the transmission network.

The reduction procedure can be summarized as the following steps:
\begin{itemize}
	\item Calculate the comprehensive criterion developed in Section \ref{sec:3}. Rank the nodes in systems using this criterion and then differentiate the retained and eliminated nodes.
	
	\item  Apply the conventional Ward equivalence method to the power grids to remove all the eliminated nodes. Remove the equivalent branches with the abnormal equivalent reactance.
	
	\item 
	Extend the retained nodes set to include all generator nodes. Conduct Ward equivalence again.
	
	\item Transfer the external generators to retained nodes using the shortest electrical path length. 
	
	\item Obtain the voltage phase angles of the nodes of the full model using the DC load flow calculation. Move and redistribute the external loads by matching the load flow results of the reduced model to those of the full model.
\end{itemize}

\section{Case studies}
To demonstrate the practicality of the proposed node selection criterion, case studies are performed on a 30-nodes system, of which the parameters are detailed in \cite{alsac1974optimal}. There are six generators and 41 branches in this system. The test system is plotted in Fig. \ref{Fig:case_eq1}, where orange color denotes the generator nodes.

Four criteria are used to rank the importance of nodes in the test system, namely, the comprehensive criterion, the extended betweenness, the net-ability, and the line congestion, and the abbreviations of these four criteria are C1, C2, C3, and C4, respectively. The end nodes of the most congested lines are considered to be important in C4. The lowest-ranked ten nodes, according to C1, C2, C3, and C4, are shown in table \ref{Tab:rank}. 

\begin{table}[htbp]
	\vspace{-0.4cm}
	\setlength{\abovecaptionskip}{-0.2cm} 
	\caption{Lowest ranked ten nodes from different criteria}\label{Tab:rank}
	\begin{center}
	\begin{tabular}{ccccccccccc}
		\hline
		\multirow{2}{*}{Criterion} & \multicolumn{10}{c}{Rank}                       \\ \cline{2-11} 
		& 21 & 22 & 23 & 24 & 25 & 26 & 27 & 28 & 29 & 30 \\ \hline
		\textbf{C1}                         & \textbf{20} & \textbf{19} & \textbf{14} & \textbf{16} & \textbf{18} & \textbf{29} & \textbf{30} & \textbf{8}  & \textbf{26} & \textbf{11} \\ 
		C2                         & 23 & 1  & 7  & 14 & 8  & 29 & 30 & 11 & 13 & 26 \\ 
		C3                         & 14 & 20 & 8  & 19 & 17 & 18 & 25 & 16 & 12 & 11 \\
		C4                         & 4 & 6 & 9  & 10 & 11 & 12 & 14 & 15 & 21 & 28 \\		
		 \hline
	\end{tabular}
	\end{center}
	\setlength{\belowcaptionskip}{-1cm}   
	\vspace{-0.4cm}
\end{table}

To further compare C1, C2, C3, and C4, all the nodes listed in table \ref{Tab:rank} are selected as eliminated nodes when building three different reduced models of the test system. The reduced models corresponding to C1, C2, C3, and C4 are denoted as eq1, eq2, eq3, and eq4. The layout of the eq1 model is depicted in Fig. \ref{Fig:case_eq1}, of which branches 10-15, 12-15, and 12-17 are the equivalent branches.
\begin{figure}[htbp] 
	\setlength{\abovecaptionskip}{-0.2cm} 
	\centering 
	\includegraphics[width=0.35\textwidth]{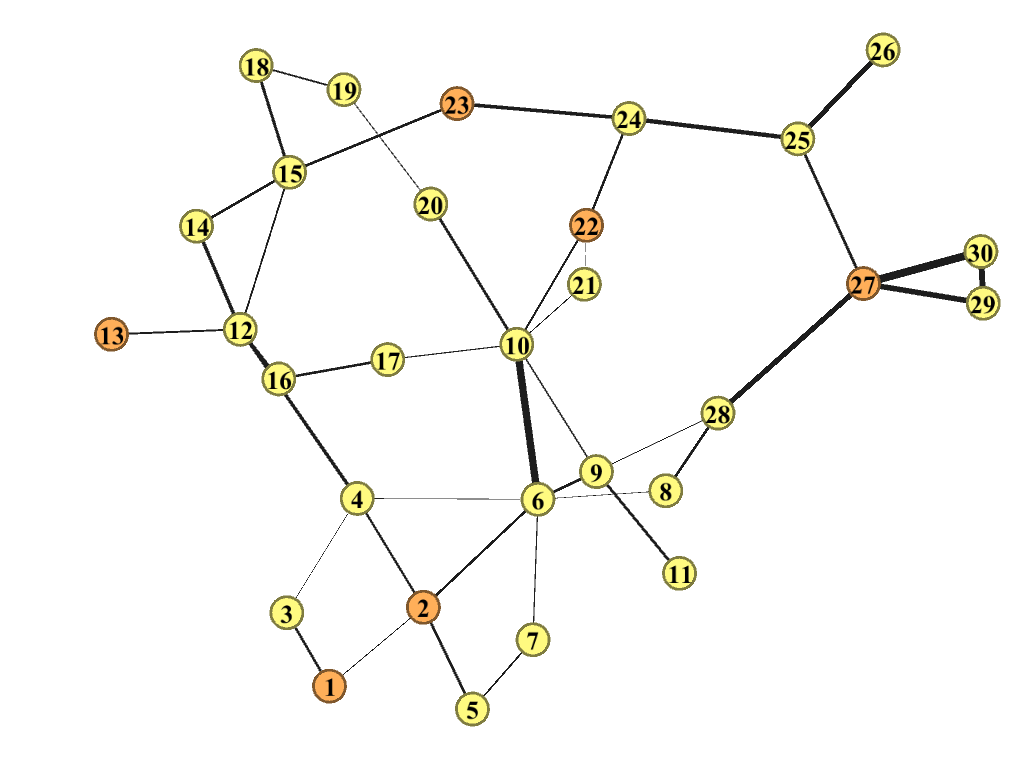} 
	\vspace{-0.8cm}
	\includegraphics[width=0.35\textwidth]{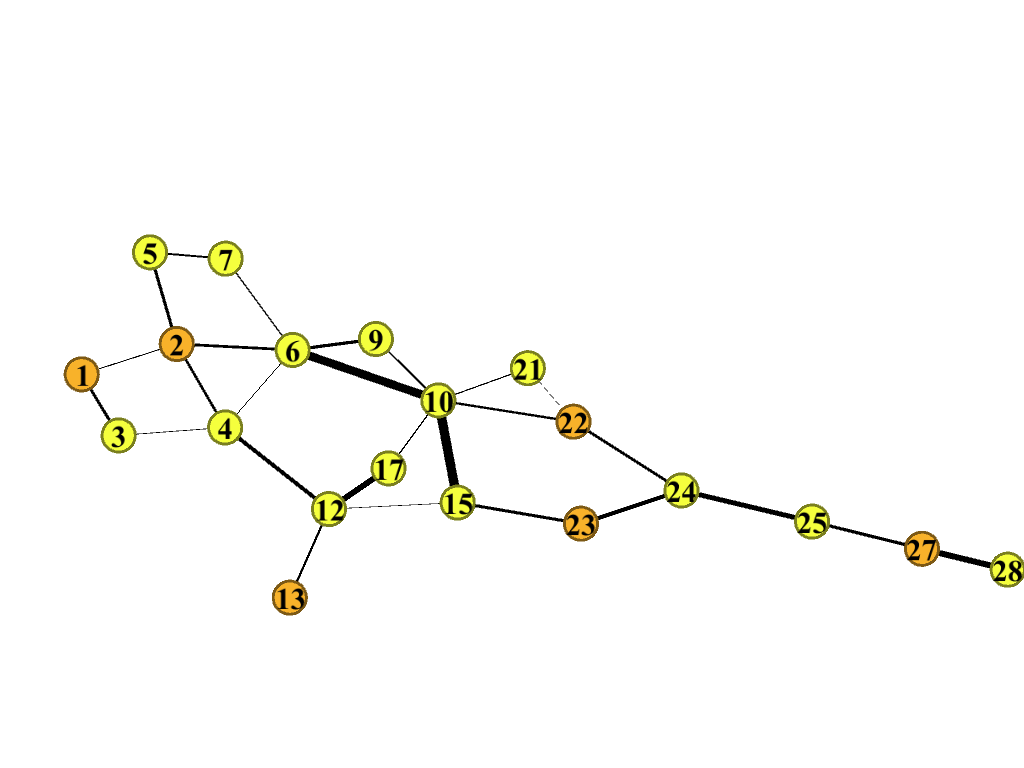} 
	\caption{Abstract layout of the full network and the reduced network 1} 
	\setlength{\belowcaptionskip}{-1cm}   
	\label{Fig:case_eq1} 
\end{figure}

The reduced models derived from the method in Section \ref{sec4} can keep the consistency of DC load flow results. As the essential metrics for evaluating the accuracy of the reduced network, the topological properties of the full and reduced networks are reported in Table \ref{tab1}, of which $E_{\text{glob}}$ is defined in (\ref{eq_abs_Eglo}), $\rho$ is the density of the network, $l_{\text{w}}$ and $l_{\text{0}}$ are the weighted and unweighted average path length, respectively, $C_{w}$ and $C_{\text{0}}$ are  the weighted and unweighted closeness centrality, respectively, $B_{w}$ and $B_{\text{0}}$ are the weighted and unweighted betweenness centrality, respectively, and $d_{\text{avg}}$ is the average degree. Herein the line weights are set as the normalized reactance of the lines. If one of the reduced models is closer to the full model on these properties compared to others, it means this reduced model can imitate the original model better. The relative error index of eq1, eq2, eq3 and eq4 corresponding to the data in Table \ref{tab1} are reported in Fig. \ref{Fig:1}. As shown in Table \ref{tab1} and Fig. \ref{Fig:1}, overall, the eq1 is the most accurate reduced model among eq1, eq2, and eq3. For example, the error of eq1 is 20.0\% of eq2, 4.2\% of eq3, and 4.0\% of eq4 on $l_{\text{w}}$, respectively. The error of eq1 is 82.0\% of eq2, 12.4\% of eq3, and 9.6\% of eq4 on $C_{\text{w}}$, respectively. The results demonstrate the superiority of the proposed criterion. The reason for the improvement of accuracy is that the proposed criterion can fully exploit the system information by considering various features of power grids simultaneously. 
\begin{table}[htbp]
\vspace{-0.4cm}
\setlength{\belowcaptionskip}{-1cm}   
\caption{Topological properties of the full and reduced networks}
\begin{center}
\begin{tabular}{cccccc}
\hline
& ori & \textbf{eq1} & eq2 & eq3 & eq4\\
\hline
$E_{\text{glob}}$ & 0.3780	& \textbf{0.4241} &	0.4274	& 0.4729 & 0.5708 \\
$\rho$ & 0.0943	& \textbf{0.1421} &	0.1421& 0.1684 & 0.2579 \\
$l_{\text{w}}$ &	1.0566	& \textbf{1.0240}	& 0.8936	& 0.2824 & 0.2358\\
$l_{\text{0}}$ &	3.3057	& \textbf{3.2053}	&  3.0632	& 2.6684 & 2.1263 \\
$C_{w}$	& 0.0353	& \textbf{0.0572}	& 0.0620	&  0.2124 & 0.2629\\
$C_{\text{0}}$ &	0.0107	& \textbf{0.0172} &	0.0178	& 0.0203 & 0.0257 \\
$B{\text{w}}$ &	39.95	& \textbf{25.45}	& 21.85	& 20.15 & 15.50\\
$B_{0}$ &	33.43	& \textbf{20.95}	& 19.60	& 15.85 & 10.70 \\
$d_{\text{avg}}$ &	2.7333&	\textbf{2.7000}&	2.7000&	3.2000 & 4.900\\
\hline
\end{tabular}
\label{tab1}
\end{center}
\vspace{-0.4cm}
\end{table}

\begin{figure}[htbp] 
	\vspace{-0.4cm}
	\centering 
	\includegraphics[width=0.45\textwidth]{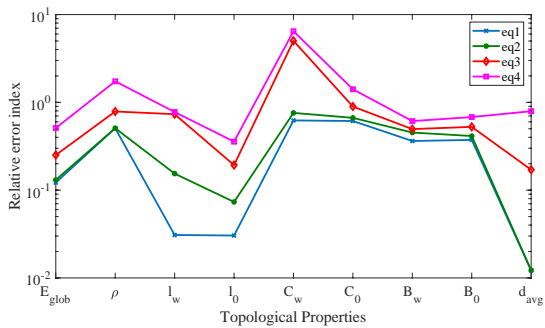} 
	\caption{The errors w.r.t topological properties} 	\label{Fig:1} 
	\setlength{\belowcaptionskip}{-1cm}   
\end{figure}

\section{Conclusion}
In this paper, a node selection criterion for network reduction in power grids was proposed. The proposed criterion can be calculated by integrating the extended betweenness and net-ability using evidential reasoning approach (ER). The case studies were conducted on a 30-nodes system. The results showed that the reduced model built upon the proposed criterion is more accurate than other models on various structural properties, which justify the advantage of the integration of the proposed criterion into the network reduction. The proposed method to derive the criterion is one of nodal feature preprocessing approaches in graph learning. Achieving network reduction via graph learning will be developed in future work.

\bibliographystyle{ieeetr}
\bibliography{conference_101719}

\end{document}